\documentstyle[12pt,epsf,amsfonts,amssymb,amsbsy]{article}
\newcommand\VV{\setbox0=\hbox{V}\hbox{\rm V\raise\ht0
  \hbox to0pt{\hss\vbox to0pt{\hbox{v}\vss}}}}
\def\slashchar#1{\setbox0=\hbox{$#1$}           
   \dimen0=\wd0                                 
   \setbox1=\hbox{/} \dimen1=\wd1               
   \ifdim\dimen0>\dimen1                        
      \rlap{\hbox to \dimen0{\hfil/\hfil}}      
      #1                                        
   \else                                        
      \rlap{\hbox to \dimen1{\hfil$#1$\hfil}}   
      /                                         
   \fi}                                         %

\begin{document}

\vspace*{1cm}

\begin{center}
{\large \bf Doubly heavy systems: decays and OPE.}\\ \vspace*{5mm}
Andrei I. Onishchenko
\end{center}
\begin{center}
{\it Institute for Theoretical and Experimental Physics,\\
Moscow, B. Cheremushkinskaja, 25, 117259
Russia \\
Fax: 7 (095) 123-65-84}.
\end{center}

\begin{abstract}
We discuss questions related to the application of OPE for the long-lived
systems with two heavy flavors. The values of
quark masses entering such calulations are constrained by use
of currently available data on the lifetimes of $B_d$ and $B_c$-mesons. The
lifetimes of doubly charmed baryons at these values of
parameters are evaluated. The necessary comments on the
difference in the results obtained previously are made.
\end{abstract}

\section{Introduction}

The main motivation for studying the weak decays of doubly heavy systems is
to test our understanding of QCD in the limit, where some
masses involved are heavy and, as a consequence, certain aspects of
bound state dynamics are simplified. In the case, when the decay of
bound system proceeds due to electro-weak interactions, the
consideration also gives us a possibility to extract
some basic properties of quark interactions at a fundamental level,
including precise determination of CKM parameters. The analysis of
decays for the doubly heavy systems already has a long history. First
of all, the decays of $J/\psi$ and $\Upsilon$ were considered.
Being composed by the quark and antiquark of the same flavor, these
mesons decay mainly trough quark-antiquark annihilation into
hadrons or lepton pair. Another particle, acquiring a lot of
attention, was the $B_c$-meson \cite{thbc}. It represents a first long-lived
particle in the family of doubly heavy systems, whose decays
proceed due to the weak interactions. The other
representatives in the family of doubly heavy hadrons are baryons
with two heavy quark, yet to be discovered experimentally. In
this paper we would like to consider lifetimes and some issues
of OPE, used in such the framework for the hadrons with two
heavy quarks, which decay due to the weak interactions.

Weak decays of the ground state of $B_c$-meson together with
semileptonic and various exclusive modes were considered in
\cite{Du,Lusig,Chang,VSR1,VSR2,KLO}. The estimates of total $B_c$-lifetime in various
quark models were done in \cite{Lusig,Chang,Quigg}. The first OPE based result
for the lifetime of $B_c$-meson was obtained in \cite{Bigi1}, where,
however, the matrix element for the dimension 5 operator $\bar Q
g\sigma\cdot G Q$ was incorrectly evaluated, which led to the overestimation of
$B_c$-lifetime\footnote{See a discussion
of this point in \cite{Beneke}.}. In addition, the leading corrections to
the spectator decays of $\bar b$ and $c$-quarks determined by the kinetic
energy, weak annihilation and Pauli interference, were not calculated
explicitly in \cite{Bigi1}. A systematical OPE-based approach to the evaluation
of $B_c$-meson lifetime was developed in \cite{Beneke}. In a short time, the
lifetime value predicted in \cite{Beneke} was experimentally confirmed by the
discovery of $B_c$-meson at FNAL by the CDF-collaboration \cite{CDFbc}.

At the next stage in studying the decays of doubly heavy systems the
approach of \cite{Beneke} was generalized to the case of baryons
with two heavy quarks in \cite{DHD1,DHD2}. A repetition of our
results at different values of parameters was done in
\cite{Guberina}. The questions related to the spectroscopy, sum rules
and production of doubly heavy baryons were considered in
\cite{DHS1,DHS2,DHSR,prod}, correspondingly.

In this work we discuss the question related to the choice of
parameters used in the OPE-based frameworks for the doubly heavy systems,
as required to obtain the theoretical predictions on the lifetimes in
agreement with the experimentally measured values.

This paper is organized as follows. In Section 2, we define
notations in the framework of OPE-calculations for the total widths of
hadrons with two heavy quarks.  In Section
3, the procedure for estimatinf the non-perturbative matrix
elements between the states with two heavy flavors is considered.
Section 4 is devoted to the numerical evaluation and discussion of parameter
choice explored for the calculation of lifetimes for the doubly charmed baryons
to be consistent with the experimentally measured value of
$B_c$-meson lifetime.  We draw conclusions in Section 5 by summarizing our
results.

\section{The OPE framework for lifetimes}

In accordance with the optical theorem, the total width
$\Gamma_{H}$ for the hadron $H$, where $H$ is the $B_c$-meson or
one of the baryons $\Xi_{cc}^{++}$, $\Xi_{cc}^{+}$  and
$\Omega_{cc}$ , has the form
\begin{equation}
\Gamma_{H} =
\frac{1}{2M_{H}}\langle H|{\cal
T} | H\rangle , \label{1}
\end{equation}
where we accept the ordinary relativistic normalization of state,
$\langle H| H\rangle  = 2EV$,
and the transition operator ${\cal T}$:
\begin{equation}
{\cal T} = \Im m\int d^4x~\{{\hat T}H_{eff}(x)H_{eff}(0)\},
\end{equation}
is determined by the effective lagrangian of weak interaction
$H_{eff}$ at the characteristic hadron energies:
\begin{equation}
H_{eff} = \frac{G_F}{2\sqrt
2}V_{q_2q_3}V_{Qq_1}^{*}[C_{+}(\mu)O_{+} + C_{-}(\mu)O_{-}] + h.c.
\end{equation}
where $$ O_{\pm} = [\bar
q_{1\alpha}\gamma_{\nu}(1-\gamma_5)Q_{\beta}][\bar
q_{2\gamma}\gamma^{\nu}(1-\gamma_5)q_{3\delta}](\delta_{\alpha\beta}\delta_{
\gamma\delta}\pm\delta_{\alpha\delta}\delta_{\gamma\beta}), $$ $$
C_+ = \left [\frac{\alpha_s(M_W)}{\alpha_s(\mu)}\right
]^{\frac{6}{33-2f}}, \quad C_- = \left
[\frac{\alpha_s(M_W)}{\alpha_s(\mu)}\right
]^{\frac{-12}{33-2f}},\\ $$ so that $f$ denotes the number of
flavors, and $Q$ marks the flavor of heavy quark ($b$ or $c$).

The quantity ${\cal T}$ in (\ref{1}) permits the Operator Product
Expansion in the inverse powers of heavy quark mass. The reason is
that, the energy release in the weak decay of either quark is
large compared to the scale of bound state dynamics, and, so, we can
expand in series over the ratio of these scales. In this way, the OPE has the
form:
\begin{equation}
{\cal T} = \sum_{i=1}^2\{C_1(\mu)\bar Q^iQ^i +
\frac{1}{m_{Q^i}^{2}}C_2(\mu)\bar Q^ig\sigma_{\mu\nu}G^{\mu\nu}Q^i
+ \frac{1}{m_{Q^i}^{3}}O(1)\}. \label{4}
\end{equation}

The leading contribution is given by the spectator decay, i.e. by
the term $\bar QQ$, which is the operator of dimension 3. The
corrections to the spectator decays of quarks are given by the operator of
dimension 5: $Q_{GQ} = \bar Q g \sigma_{\mu\nu} G^{\mu\nu} Q$ and
operators of dimension 6, $Q_{2Q2q} = \bar Q\Gamma q\bar
q\Gamma^{'}Q$, where the dominant contributions are provided by the
Pauli interference and weak scattering. For the latter, we have
\begin{eqnarray}
{\cal T}_{B_{c}^{+}} &=& {\cal T}_{35b} + {\cal T}_{35c} +
{\cal T}_{6,PI}^{(1)} + {\cal T}_{6,WA}^{(1)},\nonumber\\
{\cal T}_{\Xi_{cc}^{++}} &=& 2 {\cal T}_{35c} +
{\cal T}_{6,PI}^{(2)},\nonumber\\
{\cal T}_{\Xi_{cc}^{+}} &=& 2 {\cal T}_{35c} +
{\cal T}_{6,WS}^{(3)},\nonumber\\
{\cal T}_{\Omega_{cc}^{+}} &=& 2 {\cal T}_{35c} +
{\cal T}_{6,PI}^{(4)},\nonumber
\end{eqnarray}
where ${\cal T}_{35Q}$ denotes the contributions into the decays
of quark $Q$ by the operators with the dimensions 3 and 5, and the
forthcoming terms are the interference, scattering and annihilation of
constituents. In the explicit form we find
\begin{equation}
{\cal T}_{35b} = \Gamma_{b,spec}\bar bb -
\frac{\Gamma_{0b}}{m_b^2}[2P_{c1} + P_{c\tau 1} + K_{0b}(P_{c1} +
P_{cc1}) + K_{2b}(P_{c2} + P_{cc2})]O_{Gb}, \label{5}
\end{equation}
\begin{equation}
{\cal T}_{35c} = \Gamma_{c,spec}\bar cc -
\frac{\Gamma_{0c}}{m_c^2}[(2 + K_{0c})P_{s1} +
K_{2c}P_{s2}]O_{Gc}, \label{6}
\end{equation}
where
\begin{equation}
\Gamma_{0b} = \frac{G_F^2m_b^5}{192{\pi}^3}|V_{cb}|^2,\qquad
\Gamma_{0c} = \frac{G_F^2m_c^5}{192{\pi}^3}
\end{equation}
with $K_{0Q} = C_{-}^2 + 2C_{+}^2,~K_{2Q} = 2(C_{+}^2 - C_{-}^2)$,
and $\Gamma_{Q,spec}$ denotes the spectator width (see
\cite{Bigi2,9,10,11}):
\begin{equation}
P_{c1} = (1-y)^4,\quad P_{c2} = (1-y)^3,
\end{equation}
\begin{eqnarray}
P_{c\tau 1} &=& \sqrt{1-2(r+y)+(r-y)^2}[1 - 3(r+y) + 3(r^2+y^2) -
r^3 - y^3 -4ry + \nonumber\\ && 7ry(r+y)] +
12r^2y^2\ln\frac{(1-r-y+\sqrt{1-2(r+y)+(r-y)^2})^2}{4ry}
\end{eqnarray}
\begin{equation}
P_{cc1} = \sqrt{1-4y}(1 - 6y + 2y^2 + 12y^3)
24y^4\ln\frac{1+\sqrt{1-4y}}{1-\sqrt{1-4y}}
\end{equation}
\begin{equation}
P_{cc2} = \sqrt{1-4y}(1 + \frac{y}{2} + 3y^2) -
3y(1-2y^2)\ln\frac{1+\sqrt{1-4y}}{1-\sqrt{1-4y}}
\end{equation}
\noindent where $y = \frac{m_c^2}{m_b^2}$ and $r =
m_{\tau}^2/m_b^2$. The functions $P_{s1} (P_{s2})$ can be obtained
from $P_{c1} (P_{c2})$ by the substitution $y = m_s^2/m_c^2$. In
the $b$-quark decays, we neglect the value $m_s^2/m_b^2$ and
suppose $m_s = 0$.

The calculation of Pauli interference for the
products of heavy quark decays with the quarks in the initial
state, weak scattering and annihilation of quarks, composing the hadron,
results in:
\begin{eqnarray}
{\cal T}_{6,PI}^{(1)} &=& {\cal T}_{PI,c\bar s}^{\bar b}\\
{\cal T}_{6,WA}^{(1)} &=& {\cal T}_{WA,c\bar s} + {\cal T}_{WA,u\bar d}
+ \sum_{l} {\cal T}_{WA,\nu_l \bar l}\\
{\cal T}_{6,PI}^{(2)} &=& 2 {\cal T}_{PI,u\bar d}^c \\
{\cal T}_{6,WS}^{(3)} &=& 2 {\cal T}_{WS,cd} \\
{\cal T}_{6,PI}^{(4)} &=& 2 {\cal T}_{PI,u\bar d}^{c'} +
2 \sum_{l} {\cal T}_{PI,\nu_l \bar l}^c
\end{eqnarray}
\noindent so that
\begin{eqnarray}
2 {\cal T}_{PI,u\bar d}^c &=& -\frac{G_F^2 |\Psi^{dl} (0)|^2}{4\pi}
m_c^2(1 - \frac{m_u}{m_c})^2(m_c + m_u)\bigg\{ 10(1-z_{-})^2 -
\frac{17}{3}(1-z_{-})^2\bigg\}
\times\nonumber \\
&& \bigg((C_{+} + C_{-})^2 + \frac{1}{3}(1 - 4k^{\frac{1}{2}})(5C_{+}^2 +
C_{-}^2 -
6C_{-}C_{+})\bigg)\\
2 {\cal T}_{WS,c d} &=& \frac{3 G_F^2 |\Psi^{dl} (0)|^2}{\pi}
m_c^2(1 + \frac{m_d}{m_c})^2(m_c + m_d)(1-z_{+})^2\times \nonumber \\
&& \bigg(C_{+}^2 + C_{-}^2 + \frac{1}{3}(1 - 4k^{\frac{1}{2}})(C_{+}^2 -
C_{-}^2)\bigg)\\
2 {\cal T}_{PI,u\bar d}^{c'} &=& \frac{13 G_F^2 |\Psi^{dl} (0)|^2}{12\pi}
m_c^2(1 - \frac{m_s}{m_c})^2(m_c + m_s)\times \nonumber \\
&& \bigg((C_{+} - C_{-})^2 + \frac{1}{3}(1 - 4k^{\frac{1}{2}})(5C_{+}^2 +
C_{-}^2 +
6C_{-}C_{+})\bigg)\\
2 {\cal T}_{PI,\nu_{\tau }\bar\tau }^c &=& \frac{G_F^2 |\Psi^{dl} (0)|^2}{\pi}
m_c^2(1 - \frac{m_s}{m_c})^2(m_c + m_s)\bigg\{ 10(1-z_{-})^2 -
\frac{17}{3}(1-z_{-})^2\bigg\}
\\
{\cal T}_{PI,c\bar s}^{\bar b} &=&
\frac{G_F^2}{12\pi}|V_{cb}|^2f^2_{B_c}M_{B_c}
(m_b - m_c)^2(1-z_{-})^2(2C_{+}^2 - C_{-}^2) \\
{\cal T}_{WA,cs} &=& \frac{G_F^2}{24\pi }|V_{cb}|^2f^2_{B_c}M_{B_c}
m_c^2(1 - z_{+})^2(4C_{+}^2 + C_{-}^2 + 4C_{+}C_{-}) \\
{\cal T}_{WA,\nu\tau} &=& \frac{G_F^2}{8\pi}|V_{cb}|^2f^2_{B_c}M_{B_c}
m_{\tau}^2(1-z_{\tau})^2 \\
{\cal T}_{PI,\nu_e\bar e}^{c} &=& {\cal T}_{PI,\nu_{\mu}\bar\mu}^{c}
= {\cal T}_{PI,\nu_{\tau}\bar\tau}^{c} (z_{\tau}\to 0)
\end{eqnarray}
Here $\Psi (0)$ is the value of quark-diquark baryon wavefunction
at the origin, and $f_{B_c}$ is the leptonic constant for $B_c$-meson.
In the evolution of coefficients $C_{+}$ and $C_{-}$, we have
taken into account the threshold effects, connected to the heavy
quark masses.

In expressions (\ref{5}) and  (\ref{6}), the scale $\mu$ has been
taken approximately equal to $m_c$. In the Pauli interference
term, we suggest that the scale can be determined on the basis of
the agreement of the experimentally known difference between the
lifetimes of $\Lambda_c$, $\Xi_c^{+}$ and $\Xi_c^{0}$ with the
theoretical predictions in the framework described
above\footnote{A more expanded description is presented in
\cite{DHD1}.}. In any case, the choice of the normalization scale
leads to uncertainties in the final results.

At present, the spectator decays of heavy quarks, contributing to ${\cal
T}_{35Q}$, are known in the logarithmic approximation of QCD to
the second order \cite{12,13,14,15,16}, including the mass
corrections in the final state with the charmed quark and
$\tau$-lepton \cite{16} in the decays of $b$-quark and with the
strange quark mass for the decays of $c$-quark.  In the numerical
estimates,
we include these corrections and mass effects, but we neglect the
decay modes suppressed by the Cabibbo angle, and also the strange
quark mass effects in the $b$-decays.

Thus, the calculation of lifetimes for the doubly heavy systems under
consideraton is reduced to the problem of evaluating the matrix elements of
operators, which is the subject of next section.

\section{Hadronic matrix elements}

By use of motion equations, the matrix element of operator $\bar Q^jQ^j$ can be
expanded in series over powers of ${1}/{m_{Q^j}}$:
\begin{eqnarray}
\langle H|\bar Q^jQ^j|H\rangle
_{norm} = 1 - \frac{\langle H|\bar
Q^j[(i\boldsymbol{D})^2-(\frac{i}{2}\sigma
G)]Q^j|H\rangle_{norm}}{2m_{Q^j}^2} +
O(\frac{1}{m_{Q^j}^3}).
\end{eqnarray}
Thus, we have to estimate the matrix elements of operators from
the following list only:
\begin{eqnarray}
&& \bar Q^j(i\boldsymbol{ D})^2Q^j,\quad (\frac{i}{2})\bar
Q^j\sigma GQ^j,\quad \bar Q^j\gamma_{\alpha}(1-\gamma_5)Q^j\bar
q\gamma^{\alpha}(1-\gamma_5)q,\nonumber\\ && \bar
Q^j\gamma_{\alpha}\gamma_5Q^j\bar
q\gamma^{\alpha}(1-\gamma_5)q,\quad \bar
Q^j\gamma_{\alpha}\gamma_5Q^j\bar
Q^k\gamma^{\alpha}(1-\gamma_5)Q^k,\\ && \bar
Q^j\gamma_{\alpha}(1-\gamma_5)Q^j\bar
Q^k\gamma^{\alpha}(1-\gamma_5)Q^k.\nonumber
\end{eqnarray}
The meaning of each term in the above list was already discussed
in the previous papers on the decays of doubly heavy baryons
\cite{DHD1,DHD2} and in \cite{Beneke}, so we omit it here.

Further, employing the NRQCD expansion of operators $\bar Q Q$ and
$\bar Qg\sigma_{\mu\nu}G^{\mu\nu}Q$, we have
\begin{eqnarray}
\bar QQ &=& \Psi_Q^{\dagger}\Psi_Q -
\frac{1}{2m_Q^2}\Psi_Q^{\dagger}(i\boldsymbol{ D})^2\Psi_Q +
\frac{3}{8m_Q^4}\Psi_Q^{\dagger}(i\boldsymbol{ D})^4\Psi_Q
-\nonumber\\ &&
\frac{1}{2m_Q^2}\Psi_Q^{\dagger}g\boldsymbol{\sigma}\boldsymbol{
B}\Psi_Q - \frac{1}{4m_Q^3}\Psi_Q^{\dagger}(\boldsymbol{
D}g\boldsymbol{ E})\Psi_Q + ... \label{32}\\ \bar
Qg\sigma_{\mu\nu}G^{\mu\nu}Q &=&
-2\Psi_Q^{\dagger}g\boldsymbol{\sigma}\boldsymbol{ B}\Psi_Q -
\frac{1}{m_Q}\Psi_Q^{\dagger}(\boldsymbol{ D}g\boldsymbol{
E})\Psi_Q + ... \label{33}
\end{eqnarray}
Here the factorization at scale $\mu$ ($m_{Q} >  \mu >
m_{Q}v_{Q}$) is supposed. We have omitted the term of
$\Psi_Q^{\dagger}\boldsymbol{\sigma} (g\boldsymbol{E} \times
\boldsymbol{ D})\Psi_Q$, corresponding to the spin-orbital
interactions, which are not essential for the basic state of
hadrons under consideration. The field $\Psi_Q$ has standard
non-relativistic normalization.

Further, the phenomenological experience in the potential quark
models shows, that the kinetic energy of quarks practically does
not depend on the quark contents of system, and it is determined
by the color structure of state. So, we suppose that the kinetic
energy is equal to $T = m_bv_b^2/2 + m_cv_c^2/2$ in the $B_c$-meson and to $T =
m_dv_d^2/2 + m_lv_l^2/2$ in the case of doubly heavy baryons for the
quark-diquark system, and it is $T/2 = m_{b}v_{b}^2/2 +
m_{c}v_{c}^2/2$ in the diquark (the color factor of 1/2). Then
\begin{equation}
\frac{\langle
B_{c}|\Psi_b^{\dagger}(i\boldsymbol{D})^2\Psi_b|B_c \rangle }
{2M_{B_c}m_b^2}\simeq
v_b^2\simeq
\frac{2m_c T}{m_{b}(m_{c}+m_{b})}
\end{equation}
\begin{equation}
\frac{\langle
B_{c}|\Psi_c^{\dagger}(i\boldsymbol{D})^2\Psi_c|B_c \rangle }
{2M_{B_c}m_c^2}\simeq
v_c^2\simeq
\frac{2m_b T}{m_{c}(m_{c}+m_{b})}
\end{equation}
\begin{equation}
\frac{\langle
\Xi_{QQ'}^{\diamond}|\Psi_Q^{\dagger}(i\boldsymbol{D})^2\Psi_Q|\Xi_{QQ'}^{
\diamond} \rangle }{2M_{\Xi_{QQ'}^{\diamond}}m_Q^2}\simeq
v_Q^2\simeq
\frac{2m_qT}{(m_q+m_{Q'}+m_{Q})(m_{Q'}+m_{Q})}+\frac{m_{Q'}T}{m_Q(m_Q+m_{Q'
})}.
\end{equation}
\begin{equation}
\frac{\langle
\Xi_{QQ'}^{\diamond}|\Psi_{Q'}^{\dagger}(i\boldsymbol{D})^2\Psi_{Q'}|\Xi_{QQ'}^
{
\diamond} \rangle }{2M_{\Xi_{QQ'}^{\diamond}}m_{Q'}^2}\simeq
v_{Q'}^2\simeq
\frac{2m_qT}{(m_q+m_{Q'}+m_{Q})(m_{Q'}+m_{Q})}+\frac{m_QT}{m_{Q'}(m_Q+m_{Q'})}.
\end{equation}
Applying the quark-diquark approximation for the doubly heavy baryons and
relating the matrix element of chromomagnetic interaction of heavy quarks in
the $B_c$-meson and that of diquark with the light quark to the mass
difference between the exited and ground states $M_{H^*} - M_{H}$, we have
\begin{eqnarray}
\frac{\langle B_c|\bar cc|B_c\rangle
}{2M_{B_c}} &=& 1 -
\frac{1}{2}v_c^2 +
\frac{3}{4}\frac{M_{B_c^*}-M_{B_c}}{m_c}\bigg( 1- \frac{m_b}{2m_c}\bigg)
+ ... \nonumber\\
&\approx& 1 - 0.190 + 0.037 - 0.061 +\ldots  \\
\frac{\langle \Xi_{cc}^{\diamond}|\bar cc|\Xi_{cc}^{\diamond}\rangle
}{2M_{\Xi_{cc}^{\diamond}}} &=& 1 -
\frac{1}{2}v_c^2 -
\frac{1}{3}\frac{M_{\Xi_{cc}^{\diamond *}}-M_{\Xi_{cc}^{\diamond}}}{m_c}
- \frac{5 g^2}{18 m_c^3}|\Psi^d (0)|^2 + ... \nonumber\\
&\approx& 1 - 0.073 -0.025 - 0.009 +\ldots   \\
\frac{\langle \Omega_{cc}^{\diamond}|\bar cc|\Omega_{cc}^{\diamond}\rangle
}{2M_{\Omega_{cc}^{\diamond}}} &=& 1 -
\frac{1}{2}v_c^2 -
\frac{1}{3}\frac{M_{\Omega_{cc}^{\diamond *}}-M_{\Omega_{cc}^{\diamond}}}{m_c}
- \frac{5 g^2}{18 m_c^3}|\Psi^d (0)|^2 + ... \nonumber\\
&\approx& 1 - 0.078 -0.025 - 0.009 +\ldots
\end{eqnarray}
Our presentation here is less detailed than in previous papers
\cite{DHD1,DHD2}. However, we hope, that the interested reader can find there
all needed details. Numerically, we have assigned $T\simeq 0.4$ GeV. The values
of $|\Psi^d (0)|$ and $M_{H}$ are given in the next section.

Analogous expressions can be obtained for the matrix elements of operator
$Qg\sigma_{\mu\nu}G^{\mu\nu}Q$
\begin{eqnarray}
\frac{\langle B_c|\bar
cg\sigma_{\mu\nu}G^{\mu\nu}c|B_c\rangle
}{2M_{B_c}} &=& 3m_c(M_{B_c^*}-M_{B_c})\bigg( 1- \frac{m_b}{2m_c}\bigg)
 \approx -0.216, \\
\frac{\langle \Xi_{cc}^{\diamond}|\bar
cg\sigma_{\mu\nu}G^{\mu\nu}c|\Xi_{cc}^{\diamond}\rangle
}{2M_{\Xi_{cc}^{\diamond}}m_c^2} &=&
-\frac{4}{3}\frac{(M_{\Xi_{cc}^{\diamond *}} -
M_{\Xi_{cc}^{\diamond}})}{m_c}
- \frac{7g^2}{9m_c^3}|\Psi^d (0)|^2 \approx -0.124, \\
\langle \Omega_{QQ'}|\bar
cg\sigma_{\mu\nu}G^{\mu\nu}c|\Omega_{QQ'}\rangle &=&
\langle \Xi_{QQ'}^{\diamond}|\bar
cg\sigma_{\mu\nu}G^{\mu\nu}c|\Xi_{QQ'}^{\diamond}\rangle
\end{eqnarray}
The permutations of quark masses lead to the required expressions
for the operators of $\bar bb$ and $\bar bg\sigma_{\mu\nu}G^{\mu\nu}b$.

\section{Numerical results and Discussion}

The analysis of $B_c$-meson lifetime and lifetimes of doubly heavy
baryons considered in \cite{Beneke,DHD1,DHD2} shows a strong dependence on
quark mass values. As has been said in Introduction, the aim of this work is to
reduce this uncertainty in a way for the case of doubly heavy baryons.
For this purpose, we would like to note, that the OPE framework
used by us for the calculation of lifetimes of doubly heavy baryons is a
generalization of what was previously developed for the$B_c$-meson lifetime
\cite{Beneke}. So we can test different sets of parameters used in the
calculations of doubly heavy baryons lifetimes to the case of $B_c$-meson, for
which we already have some experimental data.

The total width of $B_c$-meson consists of the spectator $\bar b$ and $c$-quark
decays, Pauli interference and weak annihilation contributions.  The lifetime
dependence on the $b$-quark mass can be eliminated by the requirement, that for
any value of $m_c$, the $m_b$ mass value is obtained by the matching which
results in the $B_d$-meson lifetime to be equal to the experimentally
measured value $\tau_{B_d}\approx 1.55$ ps. This prescription leads to the
following approximate relation between  the heavy quark masses
\begin{equation}
m_b = m_c + 3.5 \mbox{GeV}\nonumber \label{mb-mc}
\end{equation}
The performed analysis shows, that the $B_c$-meson lifetime dependence on
the $c$-quark mass is quite large, and it is hard to eliminate the dependence
by use of some external input. For example, a quite large
$c$-quark mass or too low virtuality\footnote{At low values, the perturbation
theory cannot be justified.} are required to reproduce the absolute lifetime
and semileptonic width in the OPE-based approach. Note also, that the
application of OPE is based on the assumption about the quark-hadron duality.
The latter can be violated in the case of $D$-mesons \cite{Blok1,Blok2}, and,
thus, in this case the OPE predictions are less reliable. On the other hand, as
was noted by authors of \cite{Beneke}, there is no obvious violation
of this duality in the case of $B_c$-mesons. So, here we attempt to extract the
$c$-quark mass value by fitting the OPE result for the $B_c$-meson lifetime to
the experimentally measured value to use in the calculations of lifetimes for
the doubly heavy baryons, yet to be discovered.

In the calculations of $B_c$-meson lifetime we have used the following set of
parameters
\begin{eqnarray}
&&m_s = 0.2 \mbox{GeV}\quad |V_{cb}| =  0.04\quad M_{B_c} = 6.26
\mbox{GeV}\nonumber\\
&&M_{B_c^{*}} - M_{B_c} = 0.073 \mbox{GeV}\quad T = 0.37 \mbox{GeV}\quad
f_{B_c} = 0.5 \mbox{GeV}
\end{eqnarray}
It is just the set of parameters, implemented in the original paper on the
$B_c$ lifetime. Here, we would like to note, that the quoted value
of $f_{B_c}$ is bigger, than one obtained in the framework of QCD
sum rules \cite{Narison,Valera1,Valera2}.

In the calculation of quark spectator decays we put the
renormalization scale to $\mu = m_Q$ and in the case of nonspectator decays we
have \footnote{As in the case of $B_c$-meson, the nonspectator contributions
are not so large as compared to those of doubly charmed baryons, we have not
introduced an additional low energy logarifmic renormalization
for them.} $\mu = m_{red}$, being the reduced mass for the $B_c$-meson system.
In Table \ref{bc} we have collected the numerical values of $B_c$-meson
lifetimes and relative spectator and nonspectator contributions for the
different sets of $b$ and $c$-quark masses, satisfying relation (\ref{mb-mc}).
\begin{table}[th]
\begin{center}
\begin{tabular}{|c|c|c|c|c|c|}
\hline
Parameters, GeV & $\sum\bar b\to\bar c$, ps$^{-1}$ & $\sum c\to s$, ps$^{-1}$&
PI, ps$^{-1}$  & WA, ps$^{-1}$ & $\tau_{B_c}$, ps \\
\hline
$m_b = 5.0, m_c = 1.5, m_s =0.20$ & 0.694 & 1.148 & -0.115 & 0.193 & 0.54 \\
\hline
$m_b = 4.8, m_c = 1.35, m_s =0.15$ & 0.576 & 0.725 & -0.132 & 0.168 & 0.75 \\
\hline
$m_b = 5.1, m_c = 1.6, m_s =0.45$ & 0.635 & 1.033 & -0.101 & 0.210 & 0.55 \\
\hline
$m_b = 5.1, m_c = 1.6, m_s =0.20$ & 0.626 & 1.605 & -0.101 & 0.210 & 0.43 \\
\hline
$m_b = 5.05, m_c = 1.55, m_s =0.20$ & 0.623 & 1.323 & -0.107 & 0.201 & 0.48 \\
\hline
$m_b = 5.0, m_c = 1.5, m_s =0.15$ & 0.620 & 1.204 & -0.114 & 0.193 & 0.53 \\
\hline
\end{tabular}
\end{center}
\caption{The value of $B_c$-meson lifetime together with the spectator and
nonspectator contributions to the width at various choices of parameters.}
\label{bc}
\end{table}
Analyzing this Table, we see that, for example, the set of parameters, proposed
in \cite{Guberina} for the calculation of lifetimes for the doubly charmed
baryons, is completely inconsistent with the experimental data, when applied to
the calculation of $B_c$-lifetime. At this set, we have $\tau_{B_c} = 0.75$ ps
contrary to
\begin{equation}
\tau_{B_c}^{exp} = 0.46\pm 0.18^{stat}\pm 0.03^{syst} \mbox{ps}.
\end{equation}
The best set of parameters turns out to be $m_b = 5.05$, $m_c = 1.55$, $m_s =
0.2$ GeV. So, now we in the situation, when our set of parameters has a strong
motivation, and we may, at a confidence level, apply it to the calculation of
lifetimes for the doubly heavy baryons.

Estimating the lifetimes of doubly charmed baryons we put
\begin{eqnarray}
&&m_s = 0.2~\mbox{GeV}\quad m_s^{*} = 0.45~\mbox{GeV}\quad m_l^{*}
= 0.3~\mbox{GeV}\quad |V_{cs}| = 0.9745 \nonumber \\
&&M_{\Xi_{cc}^{++}} =  M_{\Xi_{cc}^{+}} = 3.478~\mbox{GeV}\quad
M_{\Omega_{cc}} = 3.578~\mbox{GeV} \\
&&M_{\Xi_{cc}^{\diamond *}} - M_{\Xi_{cc}^{\diamond }} =
M_{\Omega_{cc}^{*}} - M_{\Omega_{cc}} =
0.132~\mbox{GeV}\quad T = 0.4~\mbox{GeV}\quad \Psi^d (0) = 0.150
~\mbox{GeV}^{\frac{3}{2}}\nonumber
\end{eqnarray}
The numerical values of parameters, characterizing the doubly charmed
baryons, are taken from \cite{DHS1,DHS2}. $m_s^{*}$ and $m_l^{*}$
are the strange and light quark constituent masses, used for the
calculation of bound state effects in the hadronic matrix elements of
doubly charmed baryons. For the value of light quark-diquark function we assume
\begin{equation}
|\Psi^{dl} (0) |^2 = \frac{2}{3}\frac{f_D^2 M_D
k^{-\frac{4}{9}}}{12},
\end{equation}
where $f_D = 170~\mbox{MeV}$. This expression was obtained by performing the
steps similar to \cite{Rujula,Cortes} for the derivation of hyperfine splitting
in the light quark-diquark system. The factor $k^{-\frac{4}{9}}$ accounts for
the low energy logarithmic renormalization of $f_D$ constant. In the
calculation of nonspectator effects, we have accounted for the low energy
logarithmic renormalization to a hadronic scale $\mu$, where the latter was
determined from the fit of theoretical predictions for the lifetime
differences of baryons $\Lambda_c$, $\Xi_c^{+}$ and $\Xi_c^0$ to the
corresponding experimental values. Tables \ref{cc1},\ref{cc2},\ref{cc3} contain
the values of lifetimes and relative spectator and nonspectator contributions
for the doubly charmed baryons, as calculated at different sets of parameters
used previously\cite{DHD1,Guberina}, and that of obtained from the fit
of $B_c$-meson lifetime to the experimentally measured value.
\begin{table}[th]
\begin{center}
\begin{tabular}{|c|c|c|c|}
\hline
Parameters, GeV & $\sum c\to s$, ps$^{-1}$&
PI, ps$^{-1}$  & $\tau_{\Xi_{cc}^{++}}$, ps \\
\hline
$m_c = 1.35, m_s =0.15$ & 1.638 & -0.616  & 0.99 \\
\hline
$m_c = 1.6, m_s =0.45$ & 2.397 & -0.560  & 0.56 \\
\hline
$m_c = 1.55, m_s =0.2$ & 3.104 & -0.874  & 0.45 \\
\hline
\end{tabular}
\end{center}
\caption{The value of $\Xi_{cc}^{++}$ lifetime together with the spectator and
nonspectator contributions at various values of parameters.}
\label{cc1}
\end{table}
\begin{table}[th]
\begin{center}
\begin{tabular}{|c|c|c|c|}
\hline
Parameters, GeV & $\sum c\to s$, ps$^{-1}$&
WS, ps$^{-1}$  & $\tau_{\Xi_{cc}^{+}}$, ps \\
\hline
$m_c = 1.35, m_s =0.15$ & 1.638 & 1.297  & 0.34 \\
\hline
$m_c = 1.6, m_s =0.45$ & 2.397 & 2.563  & 0.20 \\
\hline
$m_c = 1.55, m_s =0.2$ & 3.104 & 1.776  & 0.20 \\
\hline
\end{tabular}
\end{center}
\caption{The value of $\Xi_{cc}^{+}$ lifetime together with the spectator and
nonspectator contributions at various values of parameters.}
\label{cc2}
\end{table}
\begin{table}[th]
\begin{center}
\begin{tabular}{|c|c|c|c|}
\hline
Parameters, GeV & $\sum c\to s$, ps$^{-1}$&
PI, ps$^{-1}$  & $\tau_{\Omega_{cc}}$, ps \\
\hline
$m_c = 1.35, m_s =0.15$ & 1.638 & 1.780  & 0.30 \\
\hline
$m_c = 1.6, m_s =0.45$ & 2.397 & 0.506  & 0.34 \\
\hline
$m_c = 1.55, m_s =0.2$ & 3.104 & 1.077  & 0.24 \\
\hline
\end{tabular}
\end{center}
\caption{The value of $\Omega_{cc}^{+}$ lifetime together with the spectator
and nonspectator contributions at various values of parameters.}
\label{cc3}
\end{table}

The lifetimes, calculated at $m_c = 1.6~\mbox{GeV}, m_s =
0.45~\mbox{GeV}$, differ from those calculated previously at these
values of parameters, because of different value for the wavefunction of
light quark-diquark system, used previously. The lifetime of
$\Xi_{cc}^{+}$ differs from that of calculated in \cite{Guberina} because of
the weak scattering contribution to the total lifetime of baryon under
consideration, which was wrongly estimated before in \cite{Guberina}.
The final comment concerns with the importance of Pauli interference in
semileptonic inclusive decays of $c$-quark, which was introduced by Voloshin.
As argued by the authors of \cite{Guberina}, this term can be valuable. So, it
is given by the following equation:
\begin{equation}
\Gamma_{SL}^{Voloshin} = \frac{G_F^2}{12\pi
}|V_{cd}|^2m_c^2(4\sqrt{k}-1)5|\Psi^{dl} (0)|^2.
\end{equation}
This term being doubly Cabbibo suppressed in the case of
$\Xi_{cc}^{+}$-baryons, does not give any sizeable contribution, when we
discuss the lifetimes of these baryons and should be taken into account only in
estimations of semileptonic branching ratios of heavy baryons. In the case of
$\Omega_{cc}$-baryon it is no longer suppressed, and, so, this term is
explicitly accounted for in our formulae.

\begin{center}
\begin{figure}[ph]
\vspace*{-1.cm} \hbox to 1.5cm
{\hspace*{3.cm}\hfil\mbox{$\tau_{\Xi_{cc}^{++}}$, ps}} \vspace*{6.5cm}
\hbox to 17.5cm {\hfil \mbox{$|\Psi^{dl} (0)|^2$, GeV$^{3}$}}
\vspace*{8.cm}\hspace*{2.5cm} \epsfxsize=12cm \epsfbox{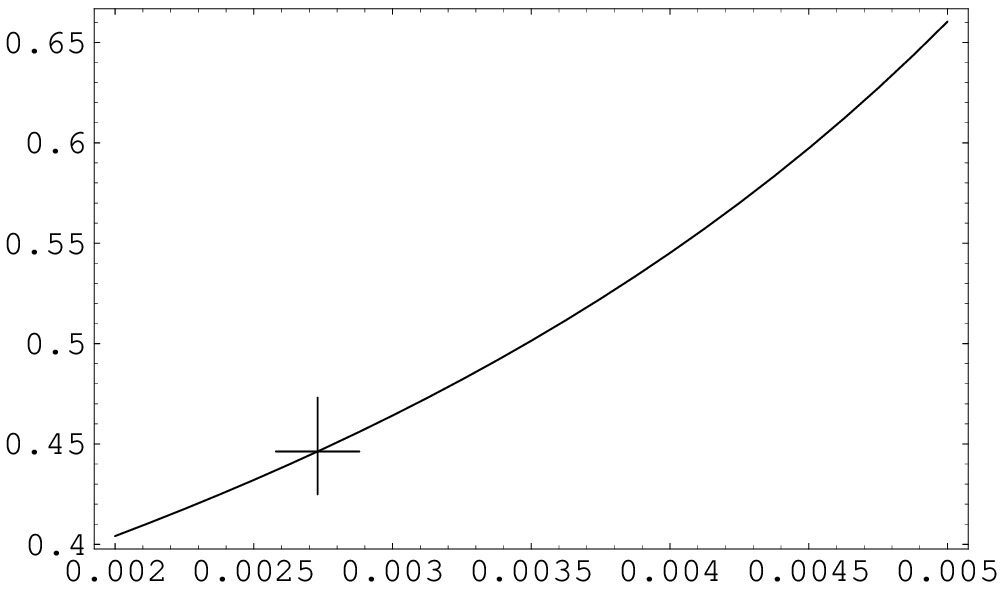}
\vspace*{-15.cm}
\caption{The dependence of $\Xi_{cc}^{++}$-baryon lifetime on the value of
wavefunction of light quark-diquark system at the origin $|\Psi^{dl} (0)|$.}
\label{ccu}
\end{figure}
\end{center}

\begin{center}
\begin{figure}[ph]
\vspace*{-1.cm} \hbox to 1.5cm
{\hspace*{3.cm}\hfil\mbox{$\tau_{\Xi_{cc}^{+}}$, ps}} \vspace*{6.5cm}
\hbox to 17.5cm {\hfil \mbox{$|\Psi^{dl} (0)|^2$, GeV$^{3}$}}
\vspace*{8.cm}\hspace*{2.5cm} \epsfxsize=12cm \epsfbox{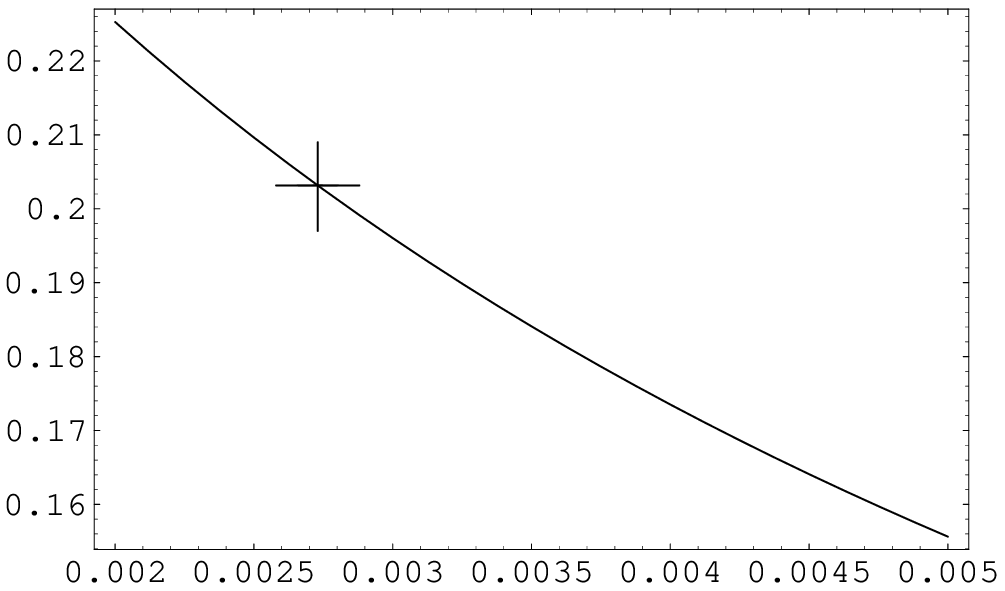}
\vspace*{-15.cm}
\caption{The dependence of $\Xi_{cc}^{+}$-baryon lifetime on the value of
wavefunction of light quark-diquark system at the origin $|\Psi^{dl} (0)|$.}
\label{ccd}
\end{figure}
\end{center}

\begin{center}
\begin{figure}[t]
\vspace*{-1.cm} \hbox to 1.5cm
{\hspace*{3.cm}\hfil\mbox{$\tau_{\Omega_{cc}}$, ps}} \vspace*{6.5cm}
\hbox to 17.5cm {\hfil \mbox{$|\Psi^{dl} (0)|^2$, GeV$^{3}$}}
\vspace*{8.cm}\hspace*{2.5cm} \epsfxsize=12cm \epsfbox{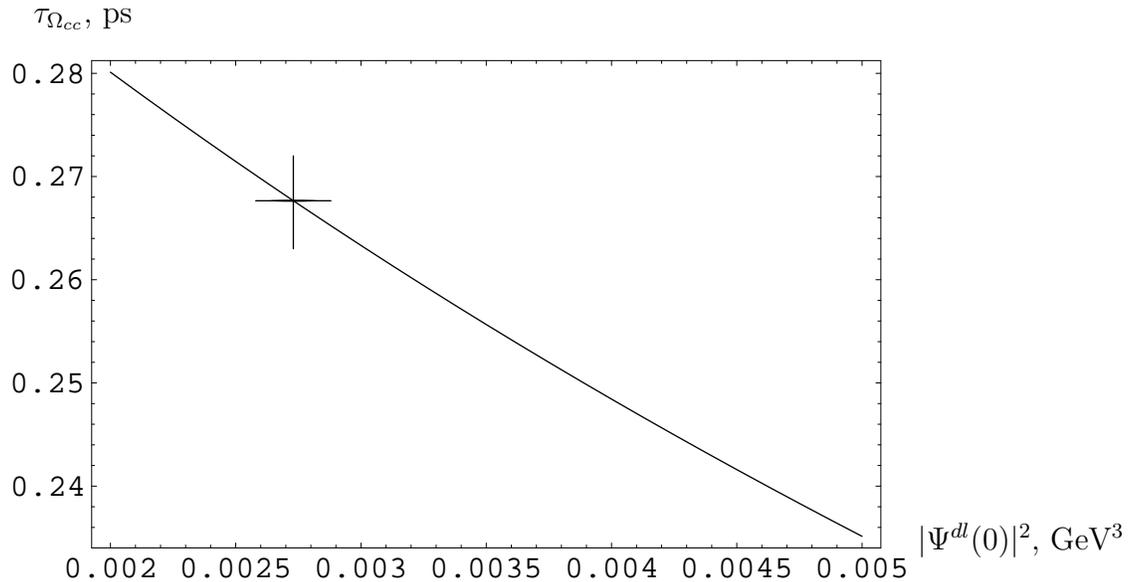}
\vspace*{-15.cm}
\caption{The dependence of $\Omega_{cc}$-baryon lifetime on the value of
wavefunction of light quark-diquark system at the origin $|\Psi^{dl} (0)|$.}
\label{ccs}
\end{figure}
\end{center}

As can be seen from the previously performed analysis
\cite{DHD1,DHD2,Guberina},
the lifetimes of doubly heavy baryons strongly depend on the value
of wavefunction for the light quark-diquark system at the origin. Also,
we can present some arguments for its determination, but in all
the cases these arguments rest on the hypothesis of quark-diquark
picture for the doubly heavy baryons. In Fig. \ref{ccu},
\ref{ccd}, \ref{ccs} we have plotted the dependence of lifetimes
for the doubly charmed baryons on this parameter. The precise value of
light quark-diquark wavefunction is under question, and, thus, the
uncertainty in its value should be included in the presented estimates
of lifetimes.

Finally we would like to comment on the theoretical errors in the
given results. They are mainly caused by the following:

1) The uncertainty in $c$-quark mass, taking into account the
experimental errors on the $B_c$-meson lifetime, can lead to
$\frac{\delta\Gamma}{\Gamma}\approx 15\%$.

2) The uncertainty in the value of light quark-diquark wave
function at the origin can lead to $\frac{\delta\Gamma}{\Gamma}\approx 30\%$.

Combining these two sources of uncertainties we get the total
error of presented estimates at the level of $45\%$.



\section{Conclusion}

In the present paper we have performed a detail investigation
of parameter influence on the lifetimes of systems with two heavy
flavors, calculated in the OPE-based approach. The fit to currently
available data on the lifetimes of $B_d$ and $B_c$-mesons, allowed us
significantly to constraint the region of heavy quark masses. We
present the numerical estimates for the lifetimes of doubly
charmed baryons and discuss the huge difference between the results obtained
previously.

This work is in part supported by the Russian Foundation for Basic
Research, grants 96-02-18216 and 96-15-96575. The work of
A.I.~Onishchenko was supported by the International Center of
Fundamental Physics in Moscow and Soros Science Foundation.
The author would like to express a gratitude to Profs. A.K.Likhoded
and V.V.Kiselev for stimulating discussions. I especially thank my wife
for a strong moral support and a help in doing the physics.

\end{document}